\documentstyle[11pt,emulateapj]{article}
\tighten

\begin{document}

\def\beginrefer{\section*{References}%
\begin{quotation}\mbox{}\par}
\def\refer#1\par{{\setlength{\parindent}{-\leftmargin}\indent#1\par}}
\def\endrefer{\end{quotation}}

\title{${\sl Chandra}$ Detects a Rapid Flare in the Gravitationally Lensed 
Mini-BALQSO RX~J0911.4+0551}

\author{G. Chartas\altaffilmark{1}, X. Dai\altaffilmark{1}, S. C. Gallagher\altaffilmark{1},
G. P. Garmire\altaffilmark{1}, M. W. Bautz\altaffilmark{2}, \\
P. L. Schechter\altaffilmark{2}, and N. D. Morgan\altaffilmark{2}}

\submitted{The Astrophysical Journal, accepted}

\altaffiltext{1}{Astronomy and Astrophysics Department, Pennsylvania State University,
University Park, PA 16802., chartas@astro.psu.edu}

\altaffiltext{2}{MIT Center for Space Research, 70 Vassar Street, Cambridge, MA, 02139.}

\begin{abstract}
The mini Broad Absorption Line (BAL) quasar RX~J0911.4+0551 was observed
with the Advanced CCD Imaging Spectrometer (ACIS) of
the {\sl Chandra X-ray Observatory} for $\sim$ 29~ks as
part of a gravitational lens (GL) survey aimed at 
measuring time-delays. Timing analysis 
of the light-curve of the lensed image A2
shows a rapid flux variation with a duration of about 2000s. 
A Kolmogorov-Smirnov test shows that the probability that a constant-intensity
source would produce the observed variability is less than $\sim$ 0.2\%.
We discuss possible origins for the observed short-term X-ray
variability.  Our gravitational lens models for the 
RX~J0911.4+0551 GL system predict a
time-delay of less than a day between images A1 and A2.
The rapid variability combined with the predicted short-time delay make
RX~J0911.4+0551 an ideal system to apply the GL method
for estimating the Hubble constant. We describe the prospects
of measuring H$_{0}$ within single X-ray observations
of GL systems with relatively short time delays.
Modeling of the spectrum of the mini-BAL quasar RX~J0911.4+0551
suggests the presence of an intrinsic absorber.
Partial covering models are slightly preferred over models
that contain absorption due to intrinsic ionized or neutral gas.

\end{abstract}


\section{INTRODUCTION}

Even before the discovery of the first gravitational 
lens system by Walsh et al. (1979) it was realized 
by Refsdal (1964) that measurements of the light travel
time difference between lensed images may provide 
an estimate of the Hubble constant. More recently it was shown 
that measurements of the time-delay can provide the absolute
distance to the intervening deflector independent 
of cosmological parameters such as H$_{0}$,
$\Omega_{m}$ and $\Omega_\Lambda$ (Narayan, 1991).
Significant observational and modeling efforts
made since then have revealed that the application of Refsdal's method 
and the uncertainties involved are more complex
than originally anticipated.

The main difficulty in measuring time-delays is that the brightness 
of each image has to be carefully monitored, continually over several periods of 
the time-delay.  Also, the quasar has to show sufficient variability 
over time-scales smaller than the time-delay.  Most attempts to 
measure time-delays until now have been made in the optical and 
radio bands.  The modest variability of quasars in these wave-bands, 
however, has made it extremely difficult to place accurate 
constraints on time-delays. 
To add to this complexity, microlensing by stars in the lensing galaxy 
may produce variability on time-scales of the order of weeks to years.
Microlensing variability of this kind does not lead to 
a time-delayed signal between the lensed images thus complicating
timing analysis. 

It has also become apparent from the extensive modeling
of several well-studied GL systems, such as Q0957+561 (Bernstein and Fisher, 1999),
that a large range of potential models for lenses can
satisfy the available observable constraints thus contributing 
to the large uncertainty in present estimates of the Hubble constant
based on Resfdal's method. X-ray observations of gravitationally 
lensed quasars may alleviate some of the complexities associated with obtaining
accurate time delays while achieving this goal in a fraction 
of the time and monitoring effort. However, X-ray observations of 
lensed quasars, have been very limited until recently due
to the relatively low spatial resolution and collecting area
of X-ray telescopes (Chartas, 2000).
Since the launch of the {\sl Chandra X-ray Observatory} on 1999 23 July,
however, the X-ray sky can now be imaged at 0.4~arcsec resolution 
and the field of gravitational lensing has suddenly become available 
to the high energy astrophysics community.

The first realization of the possibility of measuring short time
delays came with the ${\sl Chandra}$ observation
of the distant z = 2.8 quasar RX~J0911.4+0551.
The main lens of this system is a galaxy
at a redshift of 0.769 (Falco, Davis \& Stern, 1999, private communication).
A cluster of galaxies has been detected at about 38 arcsec SW from RX~J0911.4+0551
at the same mean redshift of 0.7692 $\pm$ 0.004 (Kneib et al. 2000) and is
thought to contribute to the lensing of the quasar.
The X-ray image and a deconvolved image of RX~J0911.4+0551
are presented in a companion paper by Morgan et al. (2001).

Optical spectra of RX~J0911.4+0551 obtained by Bade et al. (1997)
show an absorption trough bluewards of the C~{\sc iv} associated resonance
line that spans a velocity range of about
3200~km s$^{-1}$.  Absorption features of similar nature have been observed in
approximately 10\% of optically selected quasars
and are attributed to highly ionized gas flowing away from the central source.
Quasars observed to contain such absorption systems are commonly referred
to as Broad Absorption Line quasars (Turnshek et al. 1988; Weymann et al. 1991).
The relatively small velocity spread observed in RX~J0911.4+0551
compared to the typical range of 5,000~km s$^{-1}$ $ < $ $ v $ $ < $  25,000~km s$^{-1}$
observed in BAL quasars suggest that RX~J0911.4+0551 be classified as
a mini-BAL quasar (Turnshek 1988, Barlow, Hamann, \& Sargent 1997).

The analysis presented in the following sections focuses on the variability of the light-curves
and spectra of the lensed images of RX~0911.4+0551. The paper is structured as follows:
In section 2 we describe the timing analysis
of the ${\sl Chandra}$ observation of RX J0911.4+551,
and in section 3 we present estimates for short 
time-delays in GL systems with small image separations.
The spectral analysis of the lensed images 
is presented in section 4.
We conclude in section 5 with a summary of our results and a discussion of the prospects of constraining 
cosmological parameters by applying the short time delay method.
 
\section{TIMING ANALYSIS}

A description of the ${\sl Chandra}$ observation of RX~J0911.4+0551
is presented in a companion paper by Morgan et al. (2001).
An elementary study of the combined light-curve of 
images A1 and A2 indicates a significant variation of the X-ray flux
within a period of about 2~ks. Resolving the closest 
imaged pairs A1 and A2 (0.48~arcsec separation) 
of the GL system RX~J0911.4+0551 is just within the capabilities
of the ${\sl Chandra}$/ACIS combination.
Observations of the source PG1634-706 during
the orbital calibration phase of the mission 
demonstrated that on-axis observations, 
corrected for aspect and spacecraft dither,
can produce 50\% encircled energy radii of about 0.42~arcsec.
The observed 50\% encircled energy radius for image B of RX~J0911.4+0551 is
$\sim$ 0.38~arcsec. To search for variability in the individual images A1 and A2
we extracted light-curves from rectangular regions 
with sides of length ${\Delta}{ra}$ = 2~arcsec and ${\Delta}{dec}$ = 0.8~arcsec
centered on the centroids of
the images as determined from the deconvolution analysis
described in  Morgan et al., (2001).
The light-curves of images A1 and A2 are shown in Figure 1. 
The time resolution for this observation is
3.241sec. The light-curves in Figure 1 are binned in 1166.76s intervals,
and are corrected for background events.
Background is estimated by extracting events from
a neighboring source-free region.
The binned light-curves show a flare occurring
in image A2 and not image A1. Two consecutive data points
in the light-curve of image A2
lie at $\sim$ 3$\sigma$ above the mean value.
The duration of the X-ray flare is $\sim$ 2000s and the count-rate changed from $\sim$
5 counts per bin to $\sim$ 15 counts per bin. 
The background flares that occur at several intervals during the observation
do not coincide with the 2000s interval where we detect variability in A2.

The background rate during the X-ray flare is $\sim$ 6 $\times$ 10$^{-6}$ events s$^{-1}$ pixel$^{-1}$ 
(including only events with {\it ASCA} grades 0,2,3,4,6 and with energies within 0.5-8keV).
For the A2 source region the number
of background counts expected during the 2000s X-ray flare in A2
is $\sim$ 0.007 counts.

More appropriate and unbiased tests for unbinned data of
this nature are the Kolmogorov-Smirnov (K-S) and Kuiper tests.
In Figure 2 we show the cumulative probability distribution versus exposure time
for the light-curves of A1 and A2.
We find a significant deviation in A2, with a chance probability of about 1.8 $\times$ 10$^{-3}$
for the K-S test applied to the entire duration of the observation. Results for the Kuiper test
are similar and are presented in Table 1.
Since the X-ray flare in image A2 occurs over a brief period of $\sim$ 2000s 
we expect that the K-S probability applied over the entire observation of A2
may be a conservative value, however, we expect it to be unbiased with respect to time filtering. 
We also applied to K-S test over several narrower time intervals containing the flare.
The results are shown in Table 1. The K-S probabilities for these
time intervals range from 3.2 $\times$ 10$^{-4}$ 
to 2.3 $\times$ 10$^{-5}$ and indicate a high significance for the presence of a flare 
in image A2 but not in image A1.

An aspect error could mimic a sudden increase in the observed 
count-rate in a small extraction region. We rule
this out since an increase in X-ray flux is
still present for extraction regions that include 
all lensed images.
Also such an aspect error would produce variability in the light-curve of image A1, 
that is not observed. 
Remaining plausible causes of variability are the quasar, 
its environment and microlensing. We discuss these cases in section 4.

\section{SHORT TIME DELAYS}
The time delay between two images 
with an angular separation of ${\Delta}{\theta}$ scales 
roughly as $\sim$ ${\Delta}{\theta}^{2}$. Extrapolating from
optical and radio measurements of time-delays
we expect that time-delays in several systems with ${\Delta}{\theta}$  $ < $ 0.5~arcsec
to range between hours to days.

To obtain more accurate estimates of the short time delays for
individual systems with small angular separations
we modeled these using the $\chi^{2}$ minimization
method of Kayser et al. (1990) and Kochanek (1991).

We chose a variety of lens potentials to investigate the 
sensitivity of estimated short time delays to the mass 
distribution of the lens and to evaluate the feasibility
of observing these delays with the {\sl Chandra X-ray Observatory} and {\sl XMM-Newton}.
To account for the presence of a cluster of galaxies centered $\sim$ 38 arcsec
SW of RX~J0911.4+0551 we have included a shear component in our models.
Shear measures the anisotropic stretching of the image 
and is due to mass residing outside the beam (Weyl focusing).


The adopted two-dimensional projected potentials are,

\begin{eqnarray}
{\centering \psi_{PMXS}(\vec\theta) = b^{2}lnr  +  \frac{\gamma}{2}r^{2}\cos[2(\theta-\theta_{\gamma})]} \label{eq:phi1}\\
{\centering \psi_{ISXS}(\vec\theta) = br + \frac{\gamma}{2}r^{2}\cos[2(\theta-\theta_{\gamma})]} \label{eq:phi2}\\
{\centering \psi_{ISEP}(\vec\theta) = br + {\gamma}br~\cos[2(\theta-\theta_{\gamma})]} \label{eq:phi3}
\end{eqnarray}

\noindent
where PMXS represents a point mass with external shear model,
ISXS describes an isothermal sphere with external shear model,
and ISEP considers an isothermal elliptical potential model.
In the equations above, $r$ and $\theta$ are polar coordinates of the image
with respect to the galaxy, $\gamma$ is the shear, $\theta_{\gamma}$  
is the orientation of the shear (measured from North to West), and b is the strength of the lens.
\begin{eqnarray}
{\centering b = 4\pi\frac{D_{ls}}{c^{2}D_{os}}v^{2}_{G0}} \label{eq:b}
\end{eqnarray}

\noindent
where $v_{G0}$ is the velocity dispersion of the galaxy, D are the angular diameter
distances, and the subscripts l, s, and o refer to the lens, deflector and observer
(see Schechter et al. 1997 and references within for more details on these lens models).
The time delay between an image at a position $\vec{\theta}$ and an unlensed light path is
\begin{displaymath}
{\tau(\vec{\theta})} = {{(1+z_{d})}\over{c}}{{D_{d}D_{s}}\over{D_{ds}}}[ {{1}\over{2}}(\vec{\theta} - \vec{\beta})^{2} - \psi(\vec{\theta})] 
\end{displaymath}
where $\vec{\beta}$ is the source position.
The observable constraints are the positions of the lensed images and deflector galaxies
taken from the CfA-Arizona Space Telescope LEns Survey (CASTLES) of gravitational lenses
website {\it http://cfa-www.harvard.edu/glensdata/}. 
The best fit parameters obtained by minimizing the 
$\chi^{2}$ statistic formed between observables and their modeled 
values are listed in Table 2.
We find that in many GL systems with small-image
separations the time-delays are less than a day making them
suitable for short-time delay measurements 
with the {\sl Chandra} and {\sl XMM-Newton} observatories.
Specifically, for RX~J0911.4+0551 (Table 3) we find
that the best fit model (model ISXS in Table 3) yields a
time-delay value of 0.65$h^{-1}$days between images A1 and A2 and 
0.46$h^{-1}$days between images A1 and A3.
The detection of a flare in A2 and not in A1  
constrains the time-delay between these images 
to be greater than 23~ks, consistent with our model results.
The best fit value for the position angle of the external shear for the ISXS model
is PA = 187$\deg$. This is close to the observed position angle of PA $\sim$ 197$\deg$ 
of X-ray emission from the lensing cluster (see Morgan et al. 2001).    
Considering the additional complexity introduced by the presence
of a cluster of galaxies in the lens plane  
further detailed modeling of this system is needed.

\section{SPECTRAL ANALYSIS}

The spectra for images A1, A2, and A3 were extracted from circular regions 
centered on images with radii of 0.25~arcsec. The spectrum of image 
B was extracted from events within a 1~arcsec circular region
centered on B. The background was determined by extracting events within an
annulus centered on A1 with inner and outer radii of 10~arcsec and
15~arcsec, respectively. 
We also created a spectrum combining all lensed images 
extracting events within a circular region centered on A1 
with a radius of 10~arcsec.
Since the source spectra were in units of PHA, the data from nodes 0 and 1 are extracted 
and analyzed separately in order to account for gain differences between the amplifiers.
The parameters for each data set were linked together for the model fitting with 
the exception of the absolute normalization. 

A simple power-law model with neutral cold absorbers at z=0
was used to estimate a photon spectral index $\Gamma$
for the combined spectrum. Throughout the X-ray spectral 
analysis, the Galactic absorption is fixed at $N_{\rm H}=3.6\times10^{20}~cm^{-2}$.
We initially restricted the fit to only include events with
energies above 1.32keV ( 5~keV rest frame). By restricting the fit to these 
energies we can infer the underlying power-law continuum component
and avoid possible biasing towards harder spectra
in the event that additional absorption  by 
neutral or partially ionized gas is present.
We find a best fit value for $\Gamma$ = 1.91$^{+0.39}_{-0.34}$
typical of radio-quiet quasars at high redshifts 
(e.g., Brandt, Mathur \& Elvis 1997; Reeves \& Turner 2000),
and $\chi^2$=11.6 for 9 degrees of freedom.
Extrapolating this best fit model to lower energies reveals a
significant residual suggesting the presence of intrinsic absorption.
The resulting plot, shown in the Figure 3,
is suggestive of the possible nature of the intrinsic absorber.  The spectra
appear to recover towards the power-law continuum in the lowest energy bins, hinting at either 
partially ionized absorbing gas or partial covering of the continuum.
In contrast, a neutral, cold absorber would depress almost all flux below $\approx5$~keV, the energy where the continuum recovers.

We fit several models to try to characterize the absorbing gas.  The details 
of each fit are listed in Table 4.  
Briefly, intrinsic absorption by neutral gas at z = 2.8 with column
density $3.4\times10^{22}$~cm$^{-2}$ was preferable at the $>95$\% 
confidence level to a straight power law with only absorption fixed to the Galactic value of 
3.6 $\times$ 10$^{20}$ cm$^{-2}$ according to the $F$-test ($\Delta\chi^{2}=-5.6$).  
In addition, adding neutral intrinsic absorption
resulted in a photon index of $\Gamma=1.57^{+0.36}_{-0.29}$ which is much closer to
the value of $\Gamma$ = 1.91 for the fit to the data above 5~keV in the rest frame
than the value of $\Gamma$ = 1.25 for the fit that did not 
include intrinsic absorption.  However, the residuals still showed some systematic effects, 
in particular, a positive excess in 
the 5--11 keV (1.3--3.0 keV observed frame) range.  

We also considered models with absorption placed
at the redshift of the lensing galaxy (see Table 4).
We find no significant change in $\chi^{2}$ compared to fits
that include intrinsic absorption. We conclude that
our spectral analysis cannot constrain the redshift of the absorber.
In a latter part of this section, based on the observed properties of intrinsic absorbers
in RX~J0911.4+0551 (Bade et al. 1997),
we present physical arguments that imply
that the most likely origin of absorption in excess of the Galactic value
is the immediate quasar environment.
We, therefore, restrict the spectral analysis that follows to
models that consider intrinsic absorption.

Fitting the data with a warm absorber or a partial covering model also 
yielded acceptable fits to the data and gave reasonable photon indices(see Table 4);
the partial covering model was slightly preferred at the 63\% confidence level
over the neutral absorber model(see figure 4). 
Absorption of some kind is certainly required, however, the low signal-to-noise ratio of the 
data (the combined spectrum of all images contains a total of 425 X-ray events) 
preclude a determination of the specific nature of the absorber.  Recent X-ray gratings
analyses of the warm absorber in the Seyfert galaxy NGC~3783 (Kaspi et al. 2000)
and the absorption in the QSO IRAS 13349+2438 (Sako et al. 2001)
indicate that those absorbers are more complex than the simple one-zone photoionization 
models typically used to model CCD-resolution data.  Given that RX~J0911.4 is a 
luminous mini-BAL QSO, complex X-ray absorption is not unreasonable.

We searched for differential extinction in the spectra of the lensed
images. Because of the low signal-to-noise we kept the
spectral photon index fixed to the value of 2.0 and assumed 
Galactic and intrinsic absorption due to cold gas
at the quasar redshift. The results of these fits
are presented in Table 5. We find that 
the intrinsic column density of images
A1, A2, and B is similar and of the order 
of $\sim$ 5 $\times$ 10$^{22}$ cm$^{-2}$.
Due to the low signal-to-noise of A3 it is not possible
to constrain the column density towards the line of site
to this image within a useful bound.

The quantity $\alpha_{\rm ox} = -{{log(f_{X}/f_{opt})}\over{log(\nu_{X}/\nu_{opt})}}$, 
the spectral index of a power law defined by the flux densities 
at rest-frame 3000{\AA} and 2~keV, is a useful parameter for
measuring the X-ray power of a QSO relative to its ultraviolet continuum
emission.  A large, negative $\alpha_{\rm ox}$ indicates relatively weak
soft X-ray emission; the mean value of $\alpha_{\rm ox}$ for radio-quiet
AGN is $\approx-1.48$ (e.g., Laor et al 1997) with a typical range from
$-1.7$ to $-1.3$ (e.g., Brandt, Laor \& Wills 2000).  Weakness in soft
X-rays is plausibly explained by intrinsic X-ray absorption, which
strongly decreases the observed flux at the lowest energies.  A strong
correlation of large, negative values of $\alpha_{\rm ox}$ with the
absorption-line equivalent width of C~{\sc iv} supports this hypothesis
(Brandt, Laor, \& Wills 2000). As expected, BAL~QSOs populate the extreme
end of this correlation: they are the weakest soft X-ray sources as well
as the QSOs with the most extreme ultraviolet absorption. Recent
spectroscopic observations of two AGN with strong C~{\sc iv} absorption,
PG~1535+547 and the BAL QSO PG~2112+059, found direct evidence of
intrinsic X-ray absorption causing strongly negative values of $\alpha_{\rm
ox}$ (Gallagher et al 2001).

Given the observed C~{\sc iv} absorption-line equivalent width of
RX~J0911.4+0551, EW$_{\sc abs}=4.4\AA$ (Bade et al. 1997) and our calculated
$\alpha_{\rm ox}$ we compare this mini-BAL QSO with QSOs in the Brandt,
Laor \& Wills (2000) correlation. Specifically, we extrapolate the observed rest-frame
ultraviolet continuum from Bade et al. (1997) to 3000\AA (rest frame), and
apply the best-fitting X-ray model from our spectral analysis to obtain
the 2~keV (rest frame) flux density.
We find an observed $\alpha_{\rm ox}=-1.72$ and corrected for 
absorption, $\alpha_{\rm ox}=-1.52$, consistent with the
value for normal, unabsorbed AGN. RX~J0911.4+0551 thus fits coherently
within the correlation of Brandt, Laor \& Wills (2000).  Given that the
C~{\sc iv} absorption is undoubtedly intrinsic from the velocity spread,
$\delta V\approx3200$~km~s$^{-1}$, and redshift of the absorbers (Bade et
al. 1997), identifying the X-ray absorption as intrinsic, rather than at
the redshift of the lens, provides the most reasonable physical
explanation for the X-ray spectral modeling.

To obtain additional clues of the origin of the X-ray flare
we estimated the hardness ratio of image A2
during the X-ray flare state and the quiescent state.
For the purpose of this analysis we define as the 
X-ray flare state the time interval
between 7.2 and 9.2~ks from the beginning of the observation
and as quiescent state the remaining period of the observation.
The hardness ratio is defined as the
counts in the 1.3--8.0~keV band divided by the counts in the 0.5--1.3~keV
band.  
We find that the hardness ratio changed from 1.45 $\pm$ 0.34 in the quiescent state
to 0.67 $\pm$ 0.27 in the flare state.
If the flare resulted from a decrease in the intrinsic, absorbing column
density, then the hardness ratio would decrease as the number of counts in
the soft band increased.  The estimated hardness ratio for a
$\Gamma=1.9$ power-law with only Galactic absorption (and no intrinsic
absorption) is $\sim$ 0.6.
Correlations of spectral slope with X-ray flux have been observed
in Seyfert galaxies. In particular, a strong correlation 
of spectral slope with UV flux and broad-band 
X-ray flux have been observed in the Seyfert 1 galaxies NGC~7469,
NGC 5548 and IC4329a (Nandra, 2000; Chiang et al. 2000; Done, Madejski \& Zycki 2000).
In the Comptonization model (Haardt, Maraschi \& Ghisselini 1997;
Zdziarski, Lubinski \& Smith 1999)  an increase of the seed UV photons
leads to an increase of the Compton upscattered X-ray photons and a cooling of the disk corona,
resulting in a softened X-ray spectrum. 
Another plausible explanation for the change in hardness ratio
could arise with partial covering of the quasar by an intrinsic absorber.
Partial covering could result from lines of sight through the absorber 
combined with unobscured lines of sight off a scatterer (eg. Gallagher et. 2001).
The spectrum of the unobscured emission will be softer than
that of the absorbed emission.
If the unobscured emission is scattered into the line
of sight then the path-lengths between the X-ray source and us
will be different for the unobscured and absorbed components.
We therefore expect a delay between the arrival of these components.
A sudden flare of the quasar could thus result in a
variation of the hardness ratio.
Specifically, the hard absorbed component of the flare will 
reach us first followed by the scattered softer component. 
It is the sudden increase in the scattered component 
that we propose as the origin of the softening
of the spectrum of RX~J0911.4+0551 during the flare.

\section{DISCUSSION AND CONCLUSIONS}

Microlensing, a phenomenon induced by the random motion of stars 
in the lensing galaxies of GL systems has been proposed as the cause of 
observed wavelength-dependent variability 
in lensed images of QSO 2237+0305 (Wo{\'z}niak et al. 2000).  
The short duration and shape of the X-ray 
flare in image A2 of RX~J0911.4+0551 rule against a microlensing event. Observed
microlensing events have typical time-scales of weeks to years.
Microlensing can produce variations in the observed flux of
a lensed object when the characteristic length scale of the 
emitting region is comparable to the 
projected Einstein-ring radius,
$\zeta_{E}$ = [(4$GM/c^{2})(D_{os}D_{ls}/D_{ol}$)]$^{1/2}$, produced by a star of mass
{\it M} in the lens plane, where D represents the angular 
diameter distances, and the subscripts
{\it l, s}, and {\it o} refer to the lens, source, and observer, respectively. 
For the GL system RX~J0911+0551 
with lens and source redshifts of z$_{lens}$  = 0.769
and z$_{source}$ = 2.8 respectively, and assuming an isolated star of mass
{\it M} the Einstein-ring radius on the source plane is $\zeta_{E}$ $\sim$ 0.01 ($M/M_{\odot}$)$^{1/2}$ pc.
The expected duration of a microlensing event is
approximately equal to the time for the source to cross an Einstein-ring radius.
Assuming the transverse velocity of the source with respect to the caustic network is 
$v_{l}$ = 1000~km s$^{-1}$, the time-scale of the microlensing event is $t_{E}$ = $\zeta_{E}/v_{l}$ $\sim$ 11~yr.  

A more rapid variation, occasionally referred to as a high magnification event (HME), 
is expected to occur during a caustic crossing. Witt, Mao \& Schechter (1994) estimate the rise time of 
such an event as $t_{HME}$ $\sim$ 3.3($r_{s}/v_{l}$), where $r_{s}$ is 
the characteristic length scale for the source emission region.
For an X-ray emission region of $r_{s}$ = 10$^{-4}$~pc and for
projected source velocities ranging between 1000~km s$ ^{-1}$ and 100~km s$ ^{-1}$ the expected rise time for a
HME event is $t_{HME}$ $\sim$ 0.3 - 3~yr. 
We conclude that a caustic crossing is very unlikely to
have produced the X-ray flare.

Having ruled out instrumental effects and microlensing 
as the cause of the rapid flare in RX~J0911.4+0551 we discuss
plausible mechanisms related to the quasar and its environment.
Correlations of spectral variability of AGN with X-ray luminosity 
have been observed in a number of AGN (eg, Nandra 2000, and references therein). 
In particular, these studies indicate a strong anti-correlation between 
variability amplitude and X-ray luminosity. For example,
Seyfert galaxies with 2-10~keV luminosities ranging between   
10$^{42}$-10$^{44}$ erg s$^{-1}$ show significantly larger 
variability amplitudes than quasars with 2-10~keV luminosities ranging between
10$^{45}$-10$^{47}$ erg s$^{-1}$. 
RX~J0911.4+0551 has an unlensed 2-10~keV luminosity
of L$_{X}$ $\sim$ 5 $\times$ 10$^{44}$ erg s$^{-1}$,
where we have assumed a magnification factor of 20 (see Table 3).
Based on the observed dependence of variability amplitude with 
luminosity we expect RX~J0911.4+0551 to exhibit 
variability comparable to that seen in Seyferts. 

Assuming that the cause of the detected 
2000s X-ray flare is intrinsic to the quasar we  
place a rough estimate on the size of the emitting region
of about d $\sim$ ct = 1.5 $\times$ 10$^{-4}$~pc (corrected for time-dilation).
Note that due to the low signal-to-noise we may have just detected the peak of the
flare, thus the true duration of the flare may be slightly longer than observed.
To estimate the accuracy of measuring the time-delay between images A1 and A2
we simulated light-curves for these lensed images with
fluxes similar to those observed and added Poisson noise.
Figure 5 shows simulated light-curves for images 
A1 and A2. The intensity of the X-ray flare in A2 is normalized to that measured in
the Chandra observation of RX~J0911.4+0551. The simulated X-ray flare in A1 is 
shifted with respect to the flare in A2 by $t_{delay}$ = 75~ks.
The count-rate in the non-flare regions of the light-curves was 
set to the observed value of $\sim$ 4 $\times$ 10$^{-3}$ cnts s$^{-1}$. 
The cross-correlation function of the light-curves of images A1 and A2 
is shown in Figure 5. A time-delay of 75~ks 
is clearly recovered in the cross-correlation function.  
A Monte-Carlo realization of this simulation indicates that 
the percent fractional error in the recovered time-delay 
is $\sim$ 1.5\% (95\% confidence). We estimated
the significance of the cross-correlation peak value of 0.3
by performing a Monte-Carlo simulation for a constant source model, assuming a 
count-rate value of $\sim$ 4 $\times$ 10$^{-3}$ cnts s$^{-1}$.
We find the probability of obtaining a cross-correlation peak value of 0.3 or greater
for a constant source model to be less than $\sim$ 1 $\times$ 10$^{-6}$.
Our simple simulations, based on the observed 
rapid flare and the predicted short time-delay, 
suggest that measurements of a time-delay in the GL system RX~J0911.4+0551 
with an accuracy of about 1\% is possible within a single 
observation with the {\sl Chandra} or {\sl XMM-Newton} X-ray observatories. 
Time-delays can also be extracted from light-curves of unresolved
images via autocorrelation techniques as long
as the fluxes of the two images are similar.
Measurements of accurate short time-delays may
be possible in eight additional X-ray
bright GL systems based on the predictions of our lensing models
for these systems (see Table 2).
The potential usefulness of X-ray observations of GL systems can be appreciated if
one considers that it has taken almost 20 years 
of optical and radio monitoring to obtain a universal accepted time-delay
for GL quasar Q0957+561 to an accuracy of $\sim$ 1\%.
X-ray observations of short time-delay systems 
offer the prospect of providing the same accuracy
within 1-2 days of observing time.
In quadruple systems with measured long time-delays, 
obtained from monitoring in the optical and/or radio bands (see Fassnacht et al. 1999),
the addition of accurate short - time delays can provide
additional constraints on the gravitational potential of the lens,
thus providing tighter constraints on the Hubble constant. 

Our spectral analysis of the mini-BAL quasar
RX~J0911.4+0551 is suggestive of softening
of the X-ray spectrum during the flare state.
A plausible explanation, based on Compton upscattering models,
is that the softening of the X-ray spectral slope during the flare
is produced by cooling of the disk corona as a result of an increase in UV 
seed photons in the accretion disk.
A different explanation that is consistent 
with our spectral analysis is the scattering of an 
unabsorbed soft component in addition to the direct absorbed one.
A delay in arrival between the absorbed and 
scattered components may explain the observable variation in the
hardness ratio.
Spectral fits that include intrinsic absorption due to cold gas are preferred 
at the ($ > $ 95\%) confidence level over ones that only include Galactic absorption.
A highly ionized absorber at the redshift of RX~J0911.4+0551 is also acceptable in a 
statistical sense with best fit values of the ionization parameter and column density of 
300 erg cm s$^{-1}$ and 1 $\times$ 10$^{23}$ atoms cm$^{-2}$ respectively.
Partial covering models are slightly preferred over 
neutral and ionized absorber ones with a best fit covering
fraction of 0.7$^{+0.2}_{-0.39}$.
Spectral modeling of the individual images A1, A2, and B
suggests no evidence for significant differential extinction.
Due to the low signal to noise in image A3 it is not possible
to constrain the intrinsic column density within a useful bound. \\

We would like to thank E. Feigelson for helpful discussions
related to unbiased tests of variability.
We acknowledge financial support by NASA grant NAS 8-38252.
SGC gratefully acknowledges the support of NASA GSRP grant
NGT5-50277.

\clearpage

\small
\begin{center}
\begin{tabular}{lccccc}
\multicolumn{6}{c}{TABLE 1}\\
\multicolumn{5}{c}{TESTS OF VARIABILITY OF RX~J0911.4+0551} \\
& & & & & \\ \hline\hline
\multicolumn{1}{c} {Test} &
\multicolumn{1}{c} {Time Interval} &
\multicolumn{1}{c} {\hspace{2cm} Statistic} &
\multicolumn{1}{c} {} &
\multicolumn{1}{c} {\hspace{3cm} Chance Probability} &
\multicolumn{1}{c} {} \\
         &             &       &        &                          &  \\
         &             & A2    &   A1   &   A2                     &    A1       \\ \hline
 K-S     & 0---29.6~ks  & 0.18  & 0.05   &  1.8 $\times$ 10$^{-3}$  &   0.78  \\
Kuiper   & 0---29.6~ks  & 0.23  & 0.09   &  3.9 $\times$ 10$^{-4}$  &   0.57  \\
 K-S     & 1---28.6~ks  & 0.20  & 0.05   &  3.2 $\times$ 10$^{-4}$  &   0.81   \\
Kuiper   & 1---28.6~ks  & 0.23  & 0.10   &  5.7 $\times$ 10$^{-4}$  &   0.46   \\
 K-S     & 2---27.6~ks  & 0.23  & 0.07   &  5.2 $\times$ 10$^{-5}$  &   0.43  \\
Kuiper   & 2---27.6~ks  & 0.25  & 0.11   &  1.2 $\times$ 10$^{-4}$  &   0.27  \\
 K-S     & 3---26.6~ks  & 0.25  & 0.06   &  2.7 $\times$ 10$^{-5}$  &   0.59  \\
Kuiper   & 3---26.6~ks  & 0.26  & 0.12   &  2.1 $\times$ 10$^{-4}$  &   0.29  \\
 K-S     & 4---25.6~ks  & 0.26  & 0.08   &  2.3 $\times$ 10$^{-5}$  &   0.44  \\
Kuiper   & 4---25.6~ks  & 0.27  & 0.14   &  2.7 $\times$ 10$^{-4}$  &   0.10  \\
\hline \hline
\end{tabular}
\end{center}

\clearpage
\small
\begin{center}
\begin{tabular}{lll}
\multicolumn{3}{c}{TABLE 2}\\
\multicolumn{3}{c}{GRAVITATIONALLY LENSED SYSTEMS }\\
\multicolumn{3}{c}{WITH PREDICTED SHORT TIME DELAYS}\\
 & & \\ \hline\hline
\multicolumn{1}{c} {Object} &
\multicolumn{1}{c} {Image Separation} &
\multicolumn{1}{c} {Time-Delay} \\
                    &   arcsec     & ($h^{-1}days$) \\ \hline
RX~J0911.4+0551     & 0.47(A1-A2)  & 0.65(ISXS)\\
PG~1115+080         & 0.48(A1-A2)  & 0.065(ISXS), 0.16(SIE)\\
MG~0414+0534        & 0.42(A1-A2)  & 0.35(ISXS)\\
APM~08279+5255      & 0.15(A-C)    & 0.19(ISXS)\\
B1422+231           & 0.51(A-B)    & 0.09(ISXS), 0.37(SIE) \\
2237+0305           & 1.0(A-D)     & 0.14(SIE)\\
B0712+472           & 0.16(A-B)    &  0.05(SIE)\\
B0218+357           & 0.00137(A1-A2) & $ < $0.01(SIE)\\
1608+656            & 0.88(A-C)  & 0.85(ISXS),1.9(SIE)\\
\hline \hline
\end{tabular}
\end{center}

NOTES-\\ Time-Delays are estimates based on fitting gravitational lens models
to the observable image positions. 
We have considered isothermal sphere plus external shear (ISXS),
and singular isothermal ellipse (SIE) potentials.
The time-delays determined using SIE models
are taken from the CfA-Arizona Space Telescope LEns Survey (CASTLES) of gravitational lenses
website {\it http://cfa-www.harvard.edu/glensdata/}.

\clearpage

\scriptsize
\begin{center}
\begin{tabular}{llllllllllllll}
\multicolumn{14}{c}{TABLE 3}\\
\multicolumn{14}{c}{MODEL PARAMETERS, TIME DELAYS AND MAGNIFICATIONS FOR LENS MODELS OF RX~J0911.40551} \\
& & & & & & & & & & & & & \\ \hline\hline
\multicolumn{1}{c} {Model} &
\multicolumn{1}{c} {x } &
\multicolumn{1}{c} {y} &
\multicolumn{1}{c} {b} &
\multicolumn{1}{c} {$\gamma$}  &
\multicolumn{1}{c} {$\theta_{\gamma}$} &
\multicolumn{1}{c} {A1-A2} &
\multicolumn{1}{c} {A1-A3} &
\multicolumn{1}{c} {A1-B} &
\multicolumn{1}{c} {M$_{A1}$} &
\multicolumn{1}{c} {M$_{A2}$} &
\multicolumn{1}{c} {M$_{A3}$} &
\multicolumn{1}{c} {M$_{B}$} &
\multicolumn{1}{c} {$\chi^{2}_{\nu}/\nu$} \\ 
     &(arcsec) &(arcsec) &(arcsec) &     & (deg)  & $h^{-1}$days &$h^{-1}$days &$h^{-1}$days & & & & & \\ \hline
PMXS &-0.649   & 0.064   & 1.028   & 0.5 & 172.64 & -0.92        & 1.78        & -147.92 & -2.42 & 1.60 & -0.72 & 1.09 & 5.11/3 \\
ISXS &-0.433   & 0.047   & 1.108   & 0.31& 173.00 & -0.65        & 0.46        & -105.86 & -6.51 & 8.41 & -2.78 & 1.78 & 0.50/3 \\ 
ISEP &-0.666   &0.069    & 1.368    & 0.15& 172.23 & -0.53        & 1.19        & -149.19 & -5.85 & 4.43 & -1.88 & 1.50 & 7.88/3 \\
\hline \hline
\end{tabular}
\end{center}

NOTES-\\ PMXS represents a point mass with external shear model,
ISXS describes an isothermal sphere with external shear model,
and ISEP considers an isothermal elliptical potential model.
$\gamma$ is the shear, $\theta_{\gamma}$  
is the orientation of the shear (measured from North to West), and b is the strength of the lens.
M represents the best fit values for the magnifications
of the lensed images A1, A2, A3 and B.
{\it x} and {\it y} are the best fit positions of the source 
with respect to the lensing galaxy.
The positions of the lensed images and deflector galaxies
used to derive the lens parameters are
taken from the CASTLES website {\it http://cfa-www.harvard.edu/glensdata/}.

\clearpage
\scriptsize
\begin{center}
\begin{tabular}{lccccc}
\multicolumn{6}{c}{TABLE 4}\\
\multicolumn{5}{c}{SPECTRAL FITTING PARAMETERS FOR THE COMBINED SPECTRUM} \\
\multicolumn{5}{c}{OF IMAGES RX~J0911+0551 A1, A2, A3 and B$^{a}$} \\
& & & & & \\ \hline\hline
\multicolumn{1}{c} {Model} &
\multicolumn{1}{c} {$\Gamma$} &
\multicolumn{1}{c} {Parameter Name} &
\multicolumn{1}{c} {Value$^{b}$} &
\multicolumn{1}{c} {$\chi^2/{\nu}$} &
\multicolumn{1}{c} {$P(\chi^2/{\nu})$} \\
         &             &    &      &                       &           \\ \hline
Power Law (PL)     & $1.25^{+0.14}_{-0.15}$&... &... &28.7/20  &0.095     \\
PL, $E>5$~keV rest frame                      &$1.91^{+0.39}_{-0.34}$         &...                               &...    &11.6/9   &0.24     \\
(1.32~keV observed frame)                                   &                               &                                  &        &  &     \\
PL with intrinsic, neutral absorption                &$1.57^{+0.36}_{-0.29}$         & $N_{\rm H}(10^{22}$~cm$^{-2})$   & $3.4^{+2.6}_{-3.1}$    & 23.1/19  & 0.23     \\
PL with neutral absorption at lens                   &$1.61^{+0.28}_{-0.28}$         & $N_{\rm H}(10^{22}$~cm$^{-2})$   & $0.55^{+0.46}_{-0.36}$    & 22.6/19  & 0.25     \\
PL with intrinsic, partial-covering absorption$^{c}$ &$1.87^{+0.98}_{-0.66}$ & $N_{\rm H}(10^{22}$~cm$^{-2})$   & $19.0^{+27.9}_{-17.7}$ & 21.5/18  & 0.26     \\
                                                            &                               & Coverage fraction   & $0.71^{+0.20}_{-0.39}$      &                       &           \\
PL with intrinsic, ionized absorption$^{c}$          &$1.6^{+0.5}_{-0.4}$  &$N_{\rm H}(10^{22}$~cm$^{-2})$ &$10^{+33}_{-9}$ &23.2/18 &0.23   \\
         &             & $\xi$   & $300^{+>300}_{-300}$     &                       &           \\
PL with ionized absorption at lens$^{c}$          &$1.78^{+0.5}_{-0.4}$  &$N_{\rm H}(10^{22}$~cm$^{-2})$ &$2^{+2.4}_{-1.6}$ &21.8/18 &0.24   \\
         &             & $\xi$   & $80^{+120}_{-80}$     &                       &           \\
\hline \hline
\end{tabular}
\end{center}

NOTES-\\
$^{a}${All model fits include fixed, Galactic absorption,
$N_{\rm H}=3.72\times10^{20}$~cm$^{-2}$ (Schlegel et al. 1998).} \\
$^{b}${The errors are for 90\% confidence with all 
parameters taken to be of interest except absolute normalization.}\\
$^{c}${The parameters for this model are poorly constrained, and 
so all parameters other than $\Gamma$ have 68\% confidence errors.}

\clearpage
\normalsize

\scriptsize
\begin{center}
\begin{tabular}{lllllll}
\multicolumn{7}{c}{TABLE 5} \\
\multicolumn{7}{c}{MODEL PARAMETERS DETERMINED FROM SPECTRAL FITS TO THE } \\
\multicolumn{7}{c}{INDIVIDUAL SPECTRA OF IMAGES OF RX~J0911.4+0551 } \\
& & & & & &\\ \hline\hline
\multicolumn{1}{c} {Fit$^{a}$} &
\multicolumn{1}{c} {Image} &
\multicolumn{1}{c} {$\Gamma$} &
\multicolumn{1}{c} {$N_{H}(z=0)$} &
\multicolumn{1}{c} {$N_{H}(z=2.8)$} &
\multicolumn{1}{c} {Flux$^{b}$} &
\multicolumn{1}{c} {$\chi^{2}_{\nu}/\nu$} \\
 &            &                      & $10^{22}$$cm^{-2}$ & $10^{22}$$cm^{-2}$      &   &                \\ \hline
1(0.3-6.keV) &A1+A2+A3+B &2.0(fixed)            & 0.037(fixed)       & 5.1$_{-1.46}^{+1.94}$  & 5.2$_{-0.2}^{+0.2}$ & 1.71/21  \\
1(0.3-2.9keV)&A1         &2.0(fixed)            & 0.037(fixed)       & 4.4$_{-1.95}^{+2.65}$  &  & 1.41/14 \\
2(0.3-2.9keV)&A2         &2.0(fixed)            & 0.037(fixed)       & 5.1$_{-2.62}^{+3.88}$  &  & 1.20/8 \\
3(0.5-2.9keV)&A3         &2.0(fixed)            & 0.037(fixed)       & 1.2$_{-1.2}^{+13}$     &  & 1.6/3 \\
4(0.2-2.9keV)&A3         &2.0(fixed)            & 0.037(fixed)       & 0.2$_{-0.2}^{+2.2}$    &  & 1.54/4 \\
5(0.3-2.9)   &B          &2.0(fixed)            & 0.037(fixed)       & 3.6$_{-1.9}^{+3.1}$    &  & 0.2/2 \\ 
\hline \hline
\end{tabular}
\end{center}
\noindent
NOTES-\\
$^{a}$ The spectral fits were performed within the energy ranges listed in the parentheses. \\
$^{b}$ Flux is estimated in the 2-10keV band as is in units of
10$^{-14}$ erg s$^{-1}$ cm$^{-2}$

\clearpage

\normalsize

\beginrefer

\refer Barlow, T. A., Hamann, F., \& Sargent, W. L. W. 1977,
in ASP Conf. Ser. 28, Mass Ejection from AGN, ed. N. Arav, I. Shlosman, \& R. J. Weymann,
(San Francisco:ASP), 13 \\

\refer Bernstein, G. \& Fischer, P., 1999, \aj, 118, 14 \\

\refer Brandt, W.N., Mathur, S., \& Elvis, M. 1997, MNRAS, 285, L25 \\

\refer Brandt, W. N., Laor, A., \& Wills, B. J., 2000, \apj, 528, 637 \\

\refer {Burud}, I., {Courbin}, F., {Lidman}, C., {Jaunsen}, A.\ O.,
        {Hjorth}, J., {Ostensen}, R., {Andersen}, M.\ I., {Clasen}, J.\ W., 
        {Wucknitz}, O., {Meylan}, G., {Magain}, P., {Stabell}, R. and 
        {Refsdal}, S., 1998, \apj, 501, L5 \\

\refer Chartas, G., 2000, \apj, 531, 81 \\

\refer Chiang, J., Reynolds, C. S., Blaes, O. M., 
Nowak, M. A., Murray, N., Madejski, G., Marshall, H. L. \& 
Magdziarz, P., 2000, \apj, 528, 292 \\

\refer Done, C., Madejski, G. M., \& Zycki, P. T., 2001, \apj, in press \\

\refer Fassnacht, C. D., Pearson, T. J., Readhead, A. C. S., 
Browne, I. W. A., Koopmans, L. V. E., Myers, S. T.,
Wilkinson, P. N., 1999, \apj, 527, 498 \\

\refer {Gallagher}, S.\ C., {Brandt}, W.\ N., {Laor}, A., 
        {Elvis}, M.,{Mathur}, S., {Wills}, B.\ J. and {Iyomoto}, N., 2001, \apj, 546, 795\\

\refer Haardt, F., Maraschi, L., \& Ghisellini, G., 1997, \apj, 476, 620 \\ 

\refer Kaspi, S., Brandt, W.N., Netzer, H., Sambruna, R., Chartas, G., Garmire, G. P. \& Nousek, J.
A. 2000, ApJL, 535, L17 \\

\refer {Kayser}, R., {Surdej}, J., {Condon}, J.\ J., {Kellermann}, K.\ I., 
{Magain}, P., {Remy}, M. and {Smette}, A., 1990, \apj, 364, 15 \\

\refer Kneib, J., Cohen, J. G., \& Hjorth, J., 2000,\apj, 544, L35 \\

\refer Kochanek, C. S., 1991, \apj, 373, 354\\

\refer Laor, A., Fiore, F., Elvis, M., Wilkes, B.\ J., \& McDowell, J.\ C.\ 1997, \apj, 477, 93 \\

\refer Morgan, N. D., Chartas, G., Malm, M., Bautz, M. W., Jones, S. E., and Schechter, P. L., 2001
submitted to \apj. \\

\refer Narayan, R., 1991, \apj, 378, L5 \\

\refer Nandra, K, 2000, to appear in Advances in Space Research, astro-ph/0012448 \\

\refer Refsdal, S., 1964, MNRAS, 128, 295 \\

\refer Reeves, J. \& Turner, M. 2000, MNRS, 316, 234 \\

\refer Sako, M. et al. 2001, A\&A, in press (astro-ph/0010660) \\

\refer {Schechter}, P.\ L. , {Bailyn}, C.\ D., {Barr}, R., 
        {Barvainis}, R., {Becker}, C.\ M., {Bernstein}, G.\ M.,
        {Blakeslee}, J.\ P., {Bus}, S.\ J., {Dressler}, A., 
        {Falco}, E.\ E., {Fesen}, R.\ A., {Fischer}, P., {Gebhardt}, K.,  
        {Harmer}, D., {Hewitt}, J.\ N., {Hjorth}, J., {Hurt}, T., 
        {Jaunsen}, A.\ O., {Mateo}, M., {Mehlert}, D., {Richstone}, D.\ O., 
        {Sparke}, L.\ S., {Thorstensen}, J.\ R., {Tonry}, J.\ L., 
        {Wegner}, G., {Willmarth}, D.\ W. and {Worthey}, G., 1997, \apj, 475, L85\\

\refer Turnshek, D. A., Foltz, C. B., Grillmair, C. J.,
\& Weymann, R. J., 1988, ApJ, 325, 651 \\

\refer Walsh, D., Carswell, R. F., \& Weymann, R. J., 1979, Nature, 279, 381 \\

\refer Weymann, R. J., Morris, S. L., Foltz, C. B., \& Hewitt, P. C., 1991,
ApJ, 373, 23\\

\refer Witt, H. J., Mao, S., and Schechter, P. L., 1995, \apj, 443, 18 \\ 

\refer Wo{\'z}niak, P.\ R., {Alard}, C., {Udalski}, A., 
        {Szyma{\'n}ski}, M., {Kubiak}, M., {Pietrzy{\'n}ski}, G., and
        {Zebru{\'n}}, K., 2000, \apj, 529, 88\\

\refer Zdziarski, A., A., Lubinski, P., \& Smith, D. A., 1999, \mnras, 303, L11\\

\endrefer

\clearpage

\clearpage

\begin{figure*}[t]
\plotfiddle{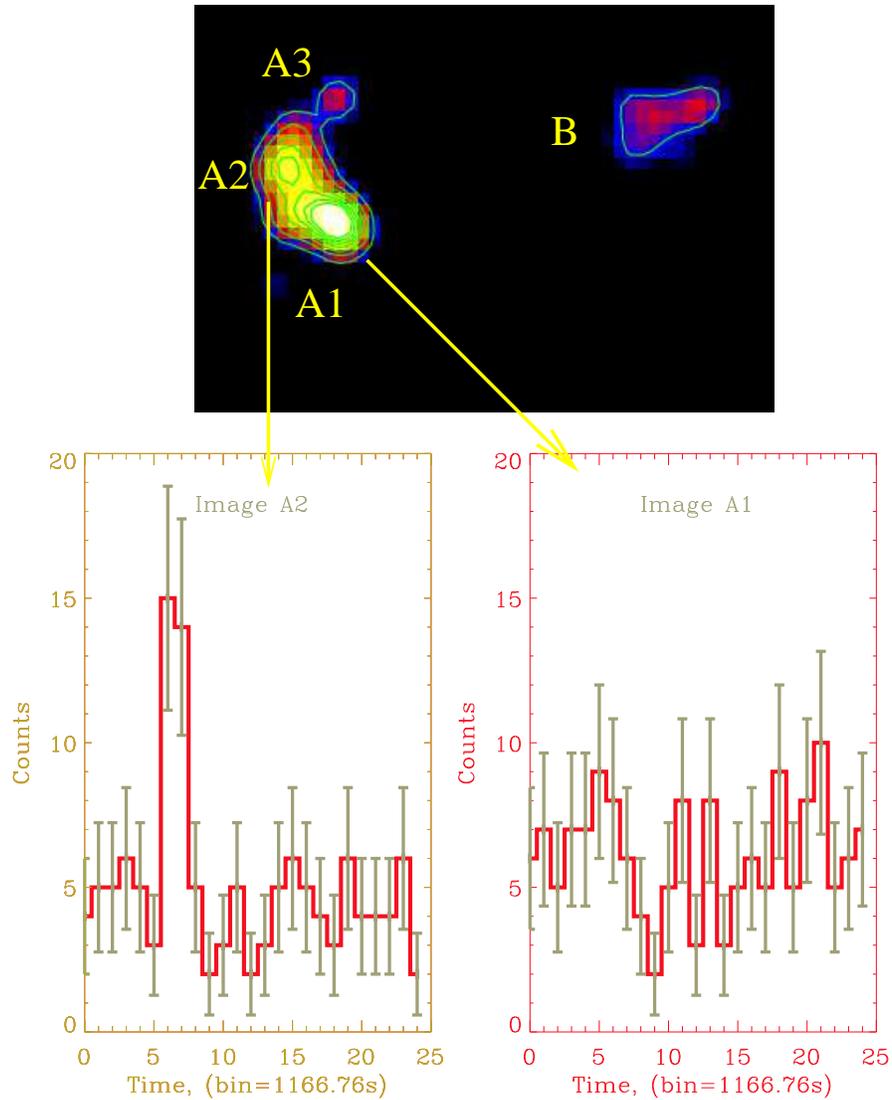}{6.5in}{0}{80.}{80.}{-250}{-100}
\protect\caption 
{\small This figure is a composite of the deconvolved X-ray image of the gravitational
lens RX~J0911.4+0551 (top panel) and the light-curves of the lensed images
A2 (left panel) and A1 (right panel). {\sl Chandra} clearly resolves the four lensed
images of the distant quasar. A rapid flare that lasted for about 2000s
was recorded in image A2 whereas image A1 does not show any variability. 
\label{fig:fig1}}
\end{figure*}

\clearpage
\begin{figure*}[t]
\plotfiddle{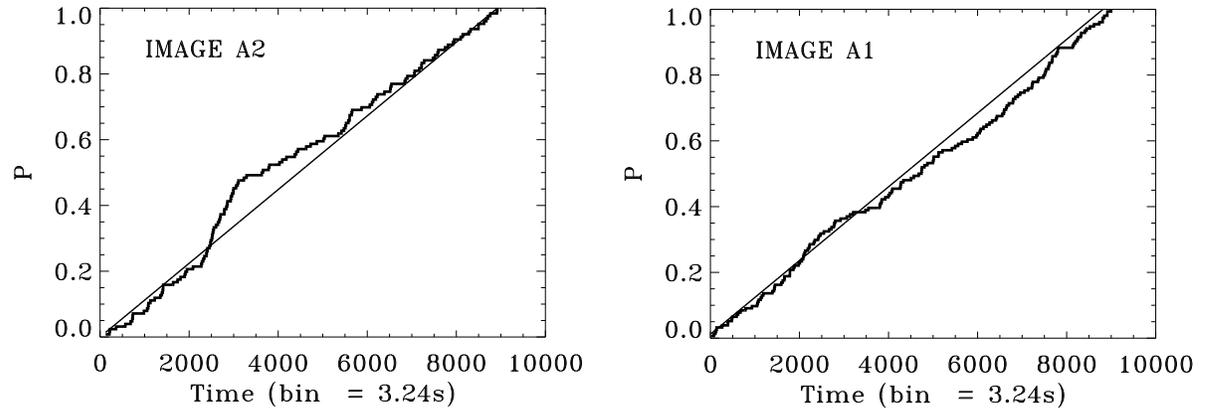}{6.in}{0}{100.}{100.}{-320}{-300}
\protect\caption
{\small Cumulative probability distribution versus exposure
number for image A2 (left panel) and image A1 (right panel)
compared to the cumulative probability distribution of a constant source.
\label{fig:fig2}}
\end{figure*}

\clearpage
\begin{figure*}[t]
\plotfiddle{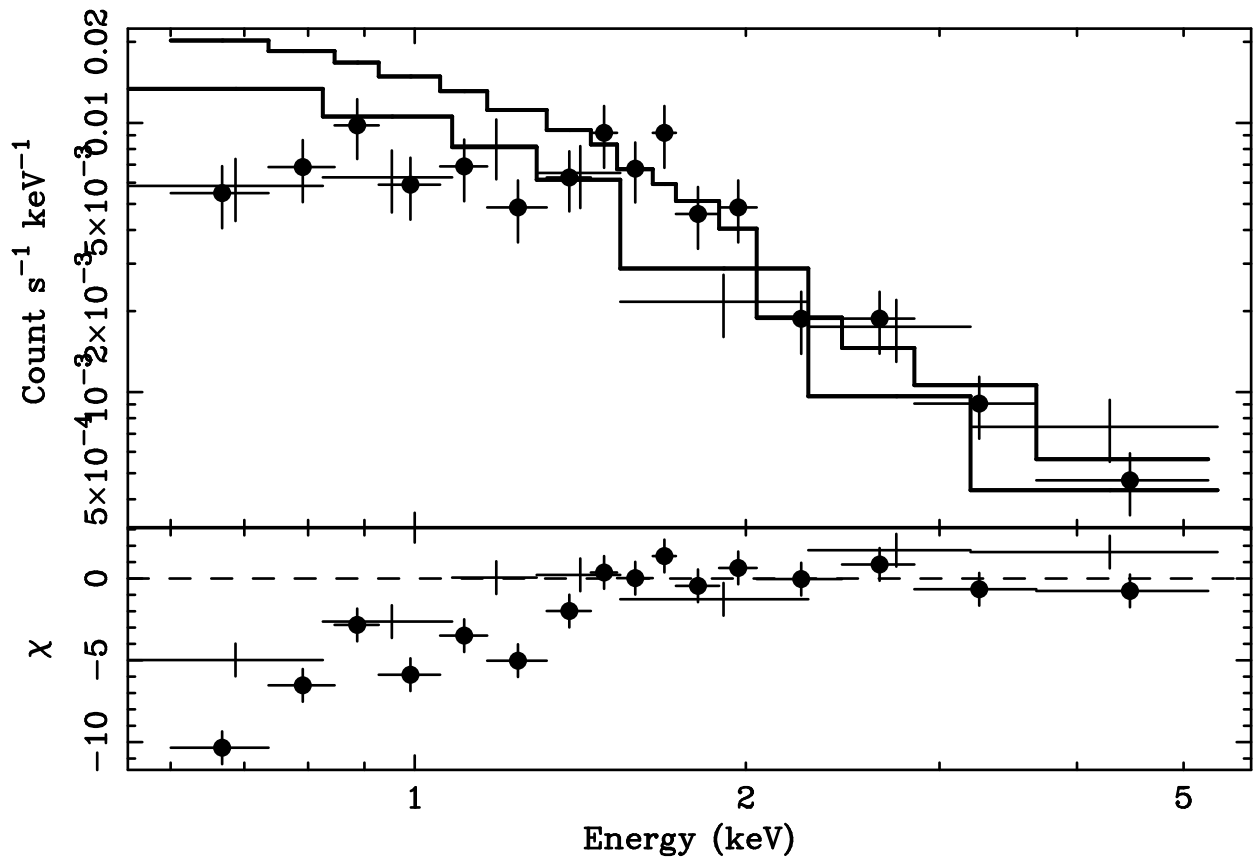}{6.5in}{-90}{70.}{70.}{-260}{470}
\protect\caption
{\small {\it Chandra} observed-frame spectra of the combined A1, A2, A3, and B
images fit with Galactic absorption and a power-law model above 1.3~keV
(5~keV rest frame) that is then extrapolated back to lower energies.
Filled circles are the data points from node 0 while plain crosses
represent the data points from node 1.  The ordinate for the lower panel,
labeled $\chi$, shows the fit residuals in terms of $\sigma$ with error
bars of size one.  Note that the flux in the lowest energy bins is not
completely extinguished. This suggests the absorber is ionized, partially
covering the continuum, or both.
\label{fig:fig3}}
\end{figure*}

\clearpage
\begin{figure*}[t]
\plotfiddle{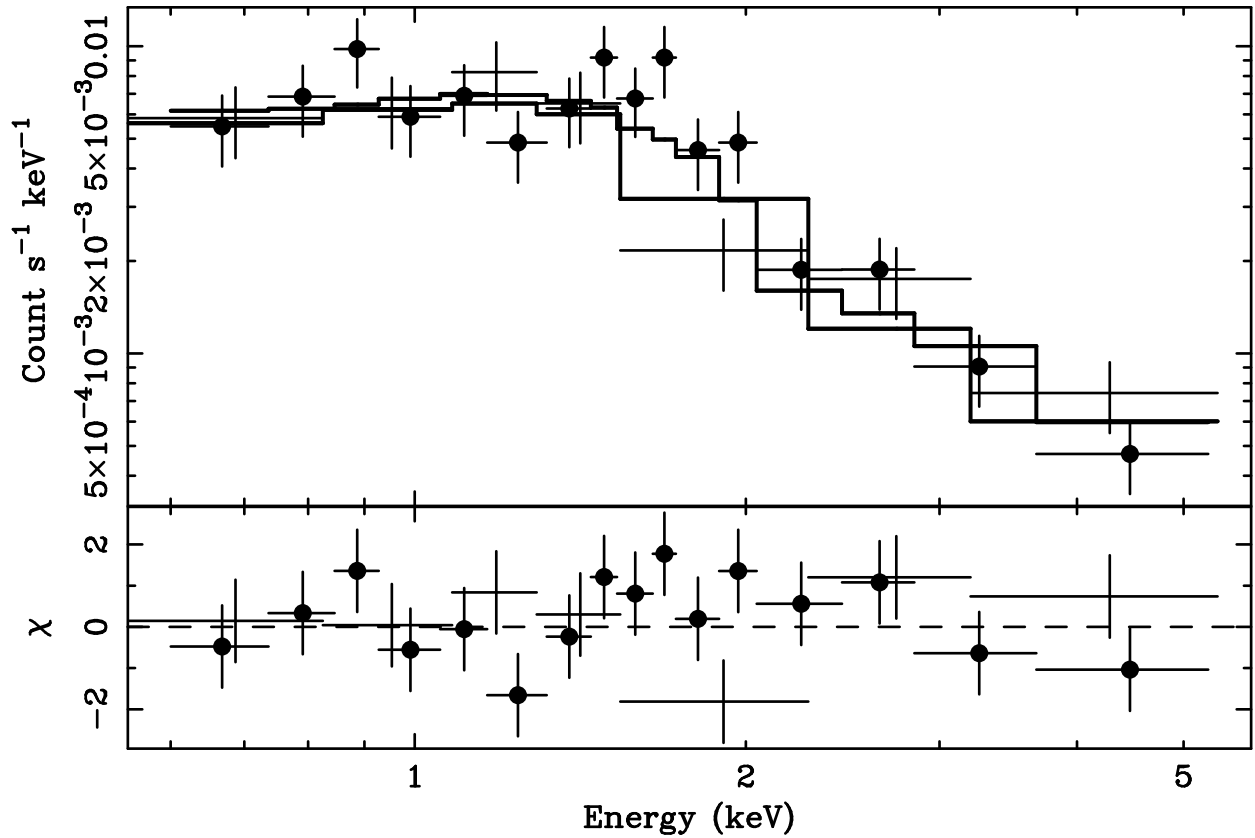}{6.5in}{-90}{70.}{70.}{-260}{470}
\protect\caption
{\small  {\it Chandra} observed-frame spectra of the combined A1, A2, A3, and B
images fit with a power-law model, partially covering intrinsic
absorption, and Galactic absorption. Filled circles are the data points
from node 0 while plain crosses represent the data points from node 1. 
The ordinate for the lower panel, labeled $\chi$, shows the fit residuals
in terms of $\sigma$ with error bars of size one.  This model provides an
acceptable fit to the data that is a statistically significant improvement
over intrinsic warm or neutral absorption. In
addition, the photon index ($\Gamma=1.9$) is consistent with the hard band
($E>5$~keV) power law (Figure~3) and the range for radio-quiet QSOs.
\label{fig:fig4}}
\end{figure*}

\clearpage
\begin{figure*}[t]
\plotfiddle{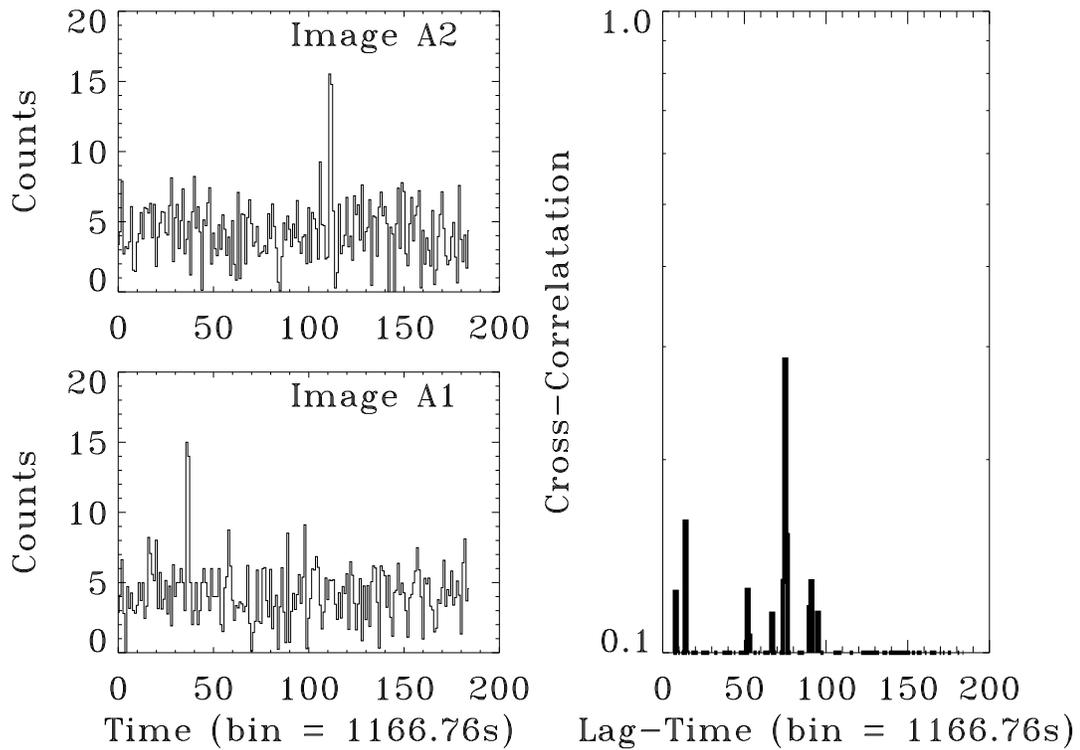}{6.5in}{90}{60.}{60.}{250}{150}
\protect\caption
{\small Simulated light-curves for images 
A1 and A2 of RX~J0911.4+0551 (left panels) with 
cross-correlation function (right panel).  
The simulated light-curves of the flare and non-flare regions 
are based on the {\sl Chandra} observations of RX~J0911.4+0551.
The input delay was set at the predicted value of 75~ks. 
The cross-correlation of light-curves A1 and A2 clearly
resolves the 75~ks time-delay.
\label{fig:fig5}}
\end{figure*}

\end{document}